\documentclass[twocolumn,english,aps,prl,superscriptaddress,showpacs]{revtex4}
\usepackage[T1]{fontenc}
\usepackage[latin9]{inputenc}
\usepackage{amstext}
\usepackage{amssymb}
\usepackage{amsmath}
\usepackage{amsthm}
\usepackage{amsfonts}
\usepackage{graphicx}
\usepackage{esint}
\usepackage{color}
\usepackage{soul}

\newcommand{\ket}[1]{\ensuremath{\left|{#1}\right\rangle}}
\newcommand{\bra}[1]{\ensuremath{\left\langle{#1}\right |}}

\newtheorem{theorem}{Theorem}

\begin{document}

\title{Uncertainty relations for characteristic functions}

\author{\L ukasz Rudnicki}

\email{rudnicki@cft.edu.pl}

\affiliation{Institute for Physics, University of Freiburg, Rheinstra\ss e 10, D-79104
Freiburg, Germany}

\affiliation{Center for Theoretical Physics, Polish Academy of Sciences, Aleja
Lotnik\'ow 32/46, PL-02-668 Warsaw, Poland}

\author{D. S. Tasca}

\affiliation{Instituto de F\'isica, Universidade Federal do Rio de Janeiro, Caixa
Postal 68528, Rio de Janeiro, RJ 21941-972, Brazil}

\author{S. P. Walborn}

\affiliation{Instituto de F\'isica, Universidade Federal do Rio de Janeiro, Caixa
Postal 68528, Rio de Janeiro, RJ 21941-972, Brazil}
\begin{abstract}
We present the uncertainty relation for the characteristic functions (ChUR)
of the quantum mechanical position and momentum probability
distributions.  This inequality is more general than the Heisenberg Uncertainty Relation, and is saturated in two extremal cases for wavefunctions described by periodic Dirac combs.  We  further discuss a broad spectrum of  applications of the ChUR, in particular, we constrain quantum optical measurements
involving general detection apertures and provide the uncertainty relation that is relevant for Loop Quantum Cosmology.  A method to measure the characteristic function directly using an auxiliary qubit is also briefly discussed. 
\end{abstract}
\maketitle

One might think that everything important has
already been said about the quantum uncertainty of conjugate
position and momentum variables, discussed for the first time
almost a century ago \cite{Heisenberg,Kennard,Robertson} in terms
of the Heisenberg Uncertainty Relation (HUR).  Even though the past few years have seen
considerable activity devoted to describing the uncertainty of non-commuting
observables in the discrete (mainly in the direction of the entropic
formulation \cite{Korzekwa1,oni,my,Coles,RPZ,Bosyk2,Bosyk3,Kaniewski,HeisEntropic} with emphasis on so-called  ``universal'' approach \cite{oni,my,oni2})
or coarse-grained \cite{HeisCoarse,OptCon,RudnickiMajCG} settings,
the most fundamental continuous position-momentum scenario appears
to be more than well understood and explored \cite{Lahti2}. For instance,
the optimal, state-independent entropic counterpart of the Heisenberg
Uncertainty Relation (HUR) was demonstrated 40 years ago \cite{BBM}, while canonically invariant uncertainty relations for higher moments
have also been derived \cite{ivan12}. In
atomic physics, where the angular momentum of electrons in an effective
central potential plays a major role, proper modifications of the
uncertainty relation for positions and momenta include the relevant
eigenvalue of the square of the angular momentum operator \cite{Dehesa},
and if the electronic state in question is not the angular momentum
eigenstate, also the variance of $\hat{L}^{2}$ \cite{RudnickiCentral}.
In the domain of Quantum Electrodynamics the ultimate Heisenberg-like
uncertainty relations have been obtained for single photon states
and the coherent states \cite{IBBURphot1,IBBURphot2}. Even the seminal
error-disturbance relation by Heisenberg, while causing problems in terms of 
rigorous interpretation, has been examined in various different
ways \cite{HeisWerner,Fujikawa,Ozawa1,OzawaExp}. 

Studies devoted to uncertainty relations are often motivated by a broad network of potential applications. In terms of quantum information, for instance, the uncertainty relations have found themselves \cite{majumdar} as important ingredients in security proofs of quantum key distribution \cite{grosshans04,branciard12}. In experimental studies within the field of quantum optics, they have been used in identification of quantum correlations such as entanglement \cite{simon00,ContEnt1,Saboia} and Einstein-Podolsky-Rosen-steering \cite{reid89,wiseman07,cavalcanti09,walborn11a,schneeloch13b}. Beyond quantum information, the uncertainty relations can play an important role in various tasks ranging from down-to-earth estimation of Hamiltonians \cite{Hamil} to pioneering experiments designed to simultaneously test Quantum Mechanics and General Relativity \cite {Magda}.

The present contribution aims to open a new chapter in the long history
of the uncertainty relations in quantum mechanics. The main subject
of our investigation is the \textit{characteristic function}, a notion well
known in classical probability theory.  The characteristic function
$\Phi\left(\lambda\right)$ related to a probability distribution
$\rho\left(x\right)$ is defined as the Fourier integral:
\begin{equation}
\Phi\left(\lambda\right)=\int_{\mathbb{R}}\! dx\, e^{i\lambda x}\rho\left(x\right).\label{definitionChar1}
\end{equation}
\par
The notion of the characteristic function acquires many interesting
features when considered on the ground of quantum mechanics. Let us assume that the probability distribution in (\ref{definitionChar1})
is related to the quantum mechanical position space.  In this case the characteristic function is equal to the average
value $\bigl\langle e^{i\lambda\hat{x}}\bigr\rangle$ of the momentum
shift operator $e^{i\lambda\hat{x}}$. Since this operator is unitary,
though non-hermitian, it is not an observable. Thus, it might not be a natural choice when thinking about uncertainty relations.
Even when calculating the variance-like quantity $\bigl\langle A^{\dagger}A\bigr\rangle$
with $A=U-\bigl\langle U\bigr\rangle$, defined for any unitary operator
$U$, one finds that the quadratic term $\bigl\langle U^{\dagger}U\bigr\rangle$
is trivially equal to $1$. The total uncertainty information is then
completely contained in the squared modulus $|\bigl\langle U\bigr\rangle|^{2}$.

Another special feature of $\Phi\left(\lambda\right)$  is related to the momentum representation.
Consider a pure state, so that $\rho\left(x\right)=|\psi\left(x\right)|^{2}$
where $\psi\left(x\right)$ is the position space wave function. The momentum wave function $\tilde{\psi}\left(p\right)$ obtained
by the Fourier transformation
\begin{equation}
\tilde{\psi}\left(p\right)=\frac{1}{\sqrt{2\pi\hbar}}\int_{\mathbb{R}}\! dx\, e^{-ipx/\hbar}\psi\left(x\right),\label{Four}
\end{equation}
naturally provides the probability distribution in momentum space
$\tilde{\rho}\left(p\right)=|\tilde{\psi}\left(p\right)|^{2}$. Definition (\ref{definitionChar1}) when rewritten in the momentum
representation gives
\begin{equation}
\Phi\left(\lambda\right)=\int_{\mathbb{R}}\! dp\,\tilde{\psi}^{*}\left(p\right)\tilde{\psi}\left(p-\hbar\lambda\right),\label{definitionChar2}
\end{equation}
which is the \textit{auto-correlation function} of the momentum wave function.
The variable $\hbar\lambda$, with units of momentum, can be thus interpreted as an
auto-correlation parameter. Obviously, when starting from the momentum distribution, an equivalent auto-correlation expression can be obtained for the position. To clearly state the distinction between the two quantum mechanical
representations let us further label by $\lambda_{x}$ and $\Phi(\lambda_{x})$
the argument and the function itself, in the case when the characteristic
function is calculated for the probability distribution $\rho\left(x\right)$.
The symbols $\lambda_{p}$ and $\tilde{\Phi}(\lambda_{p})$ shall
have the same meaning for the momentum density $\tilde{\rho}\left(p\right)$.
\par
Since the characteristic functions  $\Phi(\lambda_{x})$ and $\tilde{\Phi}(\lambda_{p})$ describe the auto-correlation  of the conjugate variables, and considering intuitively that  it should not be possible for a quantum system to be arbitrarily well localized (or more precisely, well auto-correlated) in both position and momentum variables, one would expect that there must exist some constraint on these functions, in the form of a quantum mechanical uncertainty relation. 
\par
By construction, the modulus of the characteristic function is trivially
upper-bounded by $1$. In general, however, it is hard to provide a
non-trivial, $\lambda$-dependent upper bound.  In the mathematical
literature one can find results relying on the assumptions that the
probability distribution has finite support, a given variance or is
bounded \cite{Ushakov,Rozovsky}. More elaborate studies take into
account higher moments or even the entropy \cite{Higher1,Higher2,entropyChar}
(other relevant mathematical references can be found in \cite{source,UshakovBook}).
A single restriction does not however lead to an upper bound that is sharper
than $1$, so one needs to assume more \cite{Ushakov}. Unfortunately,
both the finite support and the upper bounded value of $\rho\left(x\right)$
do not work well even for the most basic quantum-mechanical wave packets, such as
the family of Gaussians. 
For example, the variance-dependent upper bounds
(with unbounded support of the distribution) given in Ref. \cite{Ushakov} are of
the form $|\Phi(\lambda_{x})|\leq e^{-g^{2}\left(\sigma_x\right)/A_x^{2}}$ with $g\left(\sigma_x\right)=C\lambda_x/\left(2\sigma_x\left|\lambda_x\right|+\pi\right)$,
where $\sigma_x$ is the standard deviation, $A_x$ is the maximum of
the distribution and $C$ is a numerical constant.
In principle, this upper-bound 
might be useful in the current context, provided that one is able
to apply also the HUR and bound 
$\sigma_{x}\geq\hbar/(2 \sigma_{p})$.  However, it is clear that the HUR
cannot be utilized since the upper bound is a monotonically increasing function of  $\sigma_x$, and does not imply $|\Phi(\lambda_{x})|\leq e^{-g^{2}\left(\hbar/2\sigma_{p}\right)/A_{x}^{2}}$.  In addition, the optimized (with respect to  $\sigma_x$ and $\sigma_p$) sum of $e^{-g^{2}\left(\sigma_x\right)/A_x^{2}}$ and its momentum counterpart gives only a trivial bound.
\par
An easier scenario occurs when dealing with lower bounds for
the characteristic function, due to a well known fact (cf. for example
Lemma 1 of \cite{Ushakov}):
\begin{equation}
\left|\Phi\left(\lambda\right)\right|\geq\textrm{Re}\Phi\left(\lambda\right)\geq1-\frac{1}{2}\lambda^{2}\sigma^{2}.\label{lowerSD}
\end{equation}
The above inequality holds for any $\lambda$, provided that the variance
$\sigma^{2}$ is finite. However also in this case, the HUR cannot
be directly applied as it would increase the lower bound in (\ref{lowerSD}).
Let us mention that the characteristic function has been considered
in \cite{TimeEnergy}, mainly in the context of time--energy uncertainty
relations along the lines of the relation (\ref{lowerSD}). 
\par
In the
current discussion we aim to go beyond moment approximations (and
related bounds) and take into account the complete information content
stored in $e^{i\lambda x}$.  We  thus present the uncertainty relation for the position and momentum characteristic functions (ChUR):
\begin{figure}
\begin{centering}
\includegraphics[width=6cm]{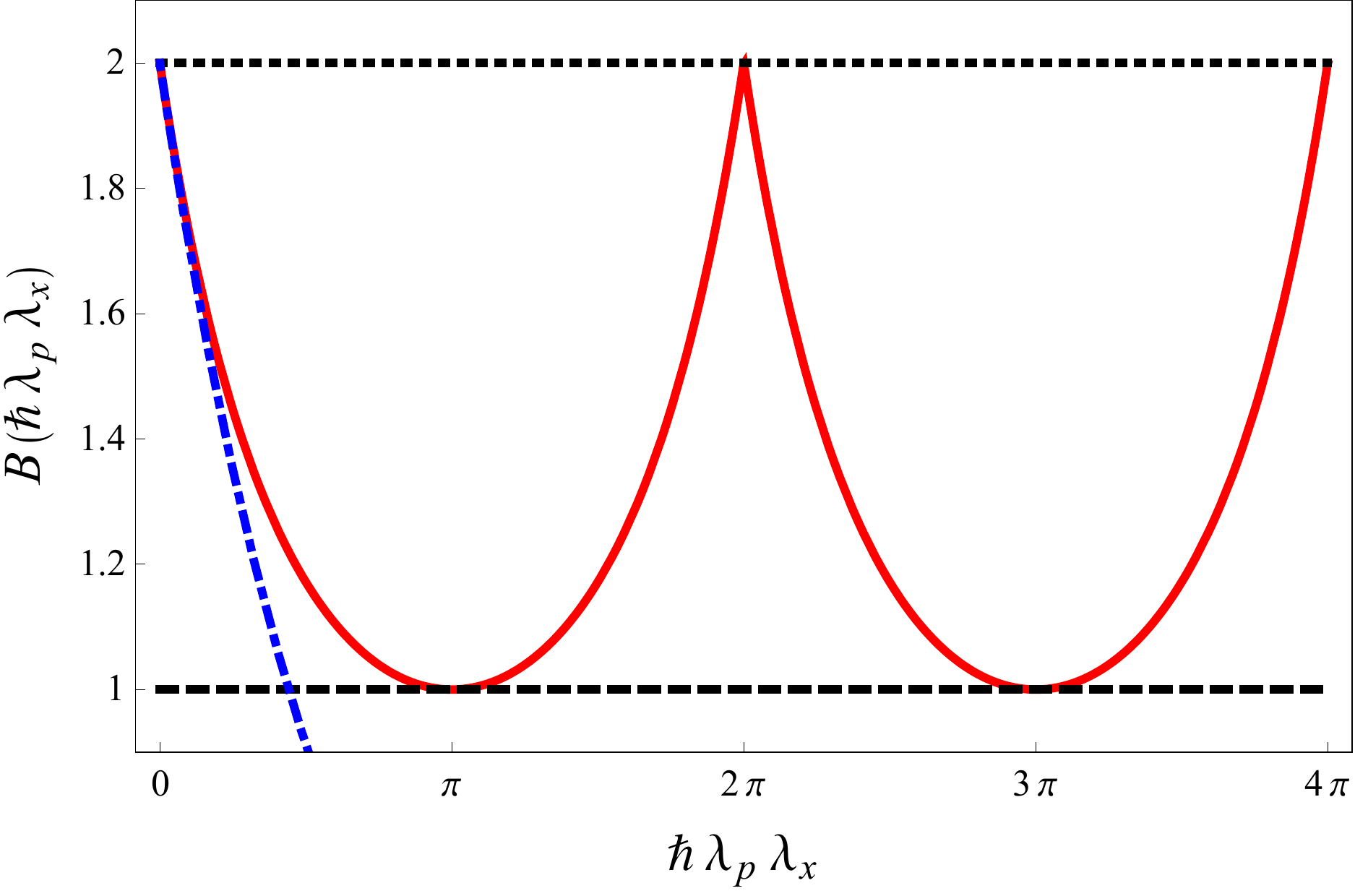}\protect\caption{(Color online). The upper bound $\mathcal{B}(\hbar\lambda_{x}\lambda_{p})$ (solid, red
line) with its upper (dotted, equal to $2$) and lower (dashed, equal to $1$) values emphasized. The blue dashed-dotted line represents the left hand side of (\ref{MajorUR1}) calculated for the Gaussian state discussed after Eq. (\ref{Eqpom}).}
\label{fig:1}
\par\end{centering}
\end{figure}
 \begin{theorem}
The sum of squared moduli of the position and momentum characteristic
functions is upper bounded:
\begin{equation}
|\Phi(\lambda_{x})|^{2}+|\tilde{\Phi}(\lambda_{p})|^{2}\leq\mathcal{B}(\hbar\lambda_{x}\lambda_{p}),\label{MajorUR1}
\end{equation}
by
\begin{equation}
\mathcal{B}(\gamma)=2\sqrt{2}\frac{\sqrt{2}-\sqrt{1-\cos\left(\gamma\right)}}{1+\cos\left(\gamma\right)}.
\end{equation}
\end{theorem}
The characteristic function is a linear functional of the probability distribution, so its modulus squared is convex. Since any quantum state can be written as a convex sum of pure states, the ChUR \eqref{MajorUR1} is satisfied by all mixed states. Moreover, the above result can be immediately generalized to the multidimensional scenario in which (\ref{definitionChar1}) is defined as a $d^N\!x$ integral with $e^{i\boldsymbol\lambda \cdot \boldsymbol{x}}$. In that case, Eq. \ref{MajorUR1} remains valid with $\lambda_x$ ($\lambda_p$) relpaced by $\boldsymbol\lambda_x$ ($\boldsymbol\lambda_p$), and consequently $\lambda_x\lambda_p$ replaced by the scalar product $\boldsymbol\lambda_x\cdot\boldsymbol\lambda_p$.
 
We will provide an outline of the proof below.  First, let us discuss the function $\mathcal{B}\left(\cdot\right)$ in inequality (\ref{MajorUR1}),   
which depends on the dimensionless parameter $\hbar\lambda_{x}\lambda_{p}$, and is plotted in Fig. \ref{fig:1}.
The upper bound is clearly periodic and varies between $1$ and $2$, the latter being the trivial value.
When $\hbar\lambda_{x}\lambda_{p}=2k\pi$ for any integer $k$ we have $\mathcal{B}(2 k \pi)=2$, and Eq. (\ref{MajorUR1}) gives no restriction on the sum of characteristic functions. This fact is a non-trivial emanation
\cite{OurMasks} of the full commutativity of spectral position and
momentum projections \cite{Projections}. In other words, the trivial
bound equal to $2$ is saturated by a distribution $\rho\left(x\right)$ that is a 
(normalized) version of the Dirac comb (wave function being the
proper limit of the sum of shifted Gaussians) with period
$2\pi/\lambda_{x}$. The characteristic function $\Phi(\lambda_{x})$ is equal to $1$
in this case. At the same time, $\tilde{\Phi}(\lambda_{p})$ is the
auto--correlation function of the comb with the related correlation
parameter $\hbar\lambda_{p}$. In this case $\hbar\lambda_{p}=2k\pi/\lambda_{x}$,
so this parameter fits the comb period, giving $\tilde{\Phi}(\lambda_{p})=1$.
In the ``opposite'' situation, when $\hbar\lambda_{x}\lambda_{p}=\left(2k+1\right)\pi$
the upper bound reaches its minimal value $1$. This case captures
the maximal non-compatibility of the position/momentum couple since
both wave functions $\psi\left(x\right)$ and $\tilde{\psi}\left(p\right)$
cannot be simultaneously well auto--correlated.  
\par
Before we prove Theorem 1 we would also like to show that the UR
(\ref{MajorUR1}) is stronger than the HUR. To this end we parametrize
$\lambda_{x}=\sqrt{a}/b$ and $\lambda_{p}=\sqrt{a}b/\hbar$, where
$a$ is a non-negative dimensionless parameter and $b$ denotes
a parameter with unit of position.  By taking the square of inequality (\ref{lowerSD})
and neglecting the positive contributions of order $\sigma^{4}$,
one immediately arrives at $|\Phi(\lambda_{x})|^{2}\geq1-\lambda_{x}^{2}\sigma_{x}^{2}$,
and the similarly for $|\tilde{\Phi}(\lambda_{p})|^{2}$. This lower bound together with the above
parametrization weakens inequality (\ref{MajorUR1}) to the form 
\begin{equation}
2-a\left(b^{-2}\sigma_{x}^{2}+\left(b/\hbar\right)^{2}\sigma_{p}^{2}\right)\leq\mathcal{B}(a).
\end{equation}
Since for $a\geq0$, $\mathcal{B}(a)=2-a+\mathcal{O}(a^{2})$, the above relation divided by $a$ implies in the limit $a\rightarrow 0$ that
\begin{equation}
1\leq b^{-2}\sigma_{x}^{2}+\left(b/\hbar\right)^{2}\sigma_{p}^{2}\label{Eqpom}.
\end{equation}
The minimum of the right hand side occurs for  $b=\sqrt{\hbar \sigma_x/\sigma_p}$ and gives the
HUR. Thus, the  ChUR \eqref{MajorUR1} is strictly stronger than the Heisenberg Uncertainty Relation. By a straightforward calculation one can check that the left hand side of (\ref{MajorUR1}) in the case of $\psi(x)$ being a Gaussian state and with all the above assignments (for $\lambda_x$, $\lambda_p$  and $b$) is equal to $2 e^{-a/2}$. In the limit  $a\rightarrow 0$, the Gaussians saturate the  bound as can be seen in Fig.~\ref{fig:1}.
\begin{proof} Let us now outline the proof of Theorem 1.  We start by taking three vectors $\left|\xi_{1}\right\rangle =\left|\Psi\right\rangle $,
$\left|\xi_{2}\right\rangle =e^{i\lambda_{x}\hat{x}}\left|\Psi\right\rangle $
and $\left|\xi_{3}\right\rangle =e^{i\lambda_{p}\hat{p}}\left|\Psi\right\rangle $
with $\left|\Psi\right\rangle $ being normalized. The positive semi-definite,
hermitian Gram matrix of this set of vectors is equal to
\begin{equation}
G=\left(\begin{array}{ccc}
1 & \Phi(\lambda_{x}) & \tilde{\Phi}(\lambda_{p})\\
\Phi^{*}(\lambda_{x}) & 1 & \Omega\\
\tilde{\Phi}^{*}(\lambda_{p}) & \Omega^{*} & 1
\end{array}\right),
\end{equation}
where $\Omega=\left\langle \Psi\right|e^{-i\lambda_{x}\hat{x}}e^{i\lambda_{p}\hat{p}}\left|\Psi\right\rangle $.
The condition of positive-semi-definiteness of $G$ leads to a single
nontrivial inequality $\det G\geq0$ which explicitly reads 
\begin{equation}
1-\Lambda-\left|\Omega\right|^{2}+\Theta+\Theta^{*}\geq0,\label{inequality1}
\end{equation}
where (we omit here the $\lambda$ arguments) $\Lambda=|\Phi|^{2}+|\tilde{\Phi}|^{2}$
and $\Theta=\Omega\Phi\tilde{\Phi}^{*}$. Consider now the parity
transformation $\lambda_{x}\mapsto-\lambda_{x}$ and $\lambda_{p}\mapsto-\lambda_{p}$.
A basic property of the characteristic function is that $\Phi(-\lambda)=\Phi^{*}(\lambda)$.
Moreover, the well known Baker\textendash Campbell\textendash Hausdorff
formula (equivalent to the Weyl commutation relations)
\begin{equation}
e^{i\lambda_{p}\hat{p}}e^{-i\lambda_{x}\hat{x}}=e^{-i\hbar\lambda_{x}\lambda_{p}}e^{-i\lambda_{x}\hat{x}}e^{i\lambda_{p}\hat{p}},
\end{equation}
provides the transformation rule $\Omega\mapsto e^{-i\hbar\lambda_{x}\lambda_{p}}\Omega^{*}$.
The terms $\Lambda$ and $\left|\Omega\right|^{2}$ present in (\ref{inequality1})
are thus invariant with respect to the above transformation, while
$\Theta\mapsto e^{-i\hbar\lambda_{x}\lambda_{p}}\Theta^{*}$. Obviously,
Eq. (\ref{inequality1}) must also hold for the transformed quantities. 

If we now take the arithmetic mean of (\ref{inequality1}) together
with its transformed counterpart, we will obtain an inequality of
almost exactly the same form as (\ref{inequality1}), with the only
difference that $\Theta$ is now multiplied by a complex constant
$Z$ (i.e. $\Theta\mapsto Z\Theta$ ) of the form $Z=\frac{1}{2}(1+e^{i\hbar\lambda_{x}\lambda_{p}})$.
In the final steps we resort to the fact that $Z\Theta\leq|Z|\left|\Theta\right|$
and the arithmetic-geometric mean inequality $\left|\Theta\right|\leq\frac{1}{2}\left|\Omega\right|\Lambda$.
The same procedure is applied to the second, conjugated term $Z^{*}\Theta^{*}$.
The above derivation leads to an inequality $1-\Lambda-\left|\Omega\right|^{2}+\left|Z\right|\left|\Omega\right|\Lambda\geq0$
which after a single rearrangement is brought to the form 
\begin{equation}
\Lambda\leq\frac{1-\left|\Omega\right|^{2}}{1-\left|Z\right|\left|\Omega\right|}.
\end{equation}
Since the parameter $0\leq|\Omega|\leq1$ can in principle assume any
value, we maximize the right hand side of the above inequality with
respect to it. The global maximum is found at $|\Omega|=(1-\sqrt{1-\left|Z\right|^{2}})/\left|Z\right|$, and
leads to the final result presented in Theorem 1, where the identity
$\left|Z\right|^{2}=\frac{1}{2}[1+\cos(\hbar\lambda_{x}\lambda_{p})]$
has been utilized. \end{proof}
\par
\begin{figure}
\begin{centering}
\includegraphics[width=6cm]{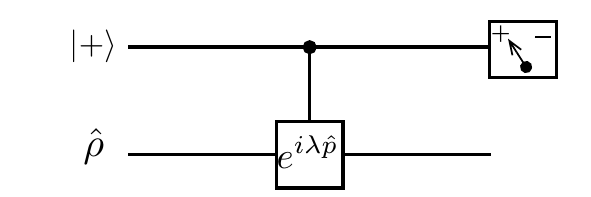}\protect\caption{Quantum circuit to measure the characteristic function of $\rho(x)$ directly.}
\label{fig:circuit}
\par\end{centering}
\end{figure}

Let us mention that if one starts the proof with an alternative choice $\left|\xi_{2}'\right\rangle =\hat{x}\left|\Psi\right\rangle $
and $\left|\xi_{3}'\right\rangle =\hat{p}\left|\Psi\right\rangle $, the counterpart of inequality (\ref{inequality1}) is equivalent to the Robertson--Schr\"odinger  uncertainty relation. Since in the current case we deal with non-hermitian displacement operators, such trivial correspondence does not occur.
\paragraph{ChUR and Quantum Optics---}Though the displacement operator does not correspond to an observable, its mean value can be measured directly in a quantum optical experiment if an ancillary qubit is used.  Consider the quantum circuit shown in Fig. \ref{fig:circuit}.  The qubit is initialized in the $\ket{+}$ state, where $\ket{\pm}=(\ket{0}\pm{\ket{1}})/\sqrt{2}$ are the eigenstates of the Pauli operator $\hat{\sigma}_x$.  The quantum system of interest is in the arbitrary state $\hat{\rho}$. The logic gate is a controlled displacement operator, defined by 
\begin{equation}
\ket{0}\bra{0} \hat{1} + \ket{1}\bra{1} e^{i \lambda_p \hat{p}}, 
\end{equation}
where the displacement is along the $x$ direction.  Detecting the qubit in the $\pm$ basis gives the 
probabilities
  \begin{equation}
P_{\pm} = \frac{1}{2}(1 \pm \langle \cos(\lambda_p \hat{p}) \rangle_{\hat{\rho}}).  
\end{equation}
Detecting the qubit in the $\hat\sigma_y$ basis, defined by eigenstates $\ket{\pm i}=(\ket{0}\pm i{\ket{1}})/\sqrt{2}$, gives
   \begin{equation}
P_{\pm i} = \frac{1}{2}(1 \mp \langle \sin(\lambda_p \hat{p}) \rangle_{\hat{\rho}}).  
\end{equation}
Then the characteristic function can be obtained by combining the measurement results, since $\langle e^{i \lambda_p \hat{p}} \rangle_{\hat{\rho}} = P_+-P_- - iP_{-i}+ iP_{+i}$.   This type of qubit-assisted measurement scheme can be realized using existing technologies in several different platforms \cite{CavityQED,hormeyll14}. 
\par

As an  application of  Theorem 1 in Quantum Optics we describe the mutual incompatibility
of measurements made in position and momentum space using detectors with arbitrary apertures. 
In our analysis we discuss a very general model, in which the detection aperture is described by an arbitrary transmittance function
$M\left(x\right)$. For single photons, for example, this detection scheme is implemented by the propagation through a general amplitude (and phase) spatial mask modelled  by $\mathcal{A}(x)=A(x)e^{i\phi(x)}$, followed by its subsequent measurement with a full multi-mode detector \cite{Tasca13c}. The  aperture function $0\leqslant A(x) \leqslant 1$ provides $M(x)=|\mathcal{A}(x)|^2\equiv A^2(x)$ while the mask phase profile  $\phi(x)$  does not affect the transmittance.
The probability $\mathcal{Q}\left(y\right)$ that the quantum particle is detected with this mask function is then 
given by 
\begin{equation}
\mathcal{Q}\left(y\right)=\int_{\mathbb{R}}\! dxM\left(x+y\right)\rho(x),
\label{convolution}
\end{equation}
where $y$ can be thought of as the location parameter that defines the mask. 
For example, if the mask is an aperture of size $\delta$ (in some
experimentally relevant units), then $M\left(x\right)$ might be chosen
to be equal $1$ for $0\leq x\leq\delta$ and $0$ elsewhere. In that
case 
(\ref{convolution}) is simply the probability of finding a quantum
particle on the interval $\left[-y,\delta-y\right]$. 

A similar construction can be done in the momentum picture. One only
needs a parameter $\kappa$ mapping the momentum variable into the
position space, so that $\kappa p$ is a position-like variable. To
have the readout function depending on the position-like variable
we define the counterpart of (\ref{convolution}) as follows
\begin{equation}
\mathcal{P}\left(y\right)=\int_{\mathbb{R}}\! dpM\left(\kappa p+y\right)\tilde{\rho}(p).
\end{equation}

By calculating the ordinary Fourier transform {[}like in Eq. (\ref{Four}),
but with $\hbar=1$ and the conjugate parameter $\lambda$ in units
of inverted position{]} of both detection probability functions we obtain $\tilde{\mathcal{Q}}(\lambda)=\tilde{M}(\lambda)\Phi(\lambda)$
and $\tilde{\mathcal{P}}(\lambda)=\tilde{M}(\lambda)\tilde{\Phi}(\kappa\lambda)$.
We are thus in position to propose a general uncertainty relation for
detection masks:
\begin{equation}
\int_{\mathbb{R}}\! dy\left(|\mathcal{Q}(y)|^{2}+|\mathcal{P}(y)|^{2}\right)\leq\int_{\mathbb{R}}\! d\lambda|\tilde{M}(\lambda)|^{2}\mathcal{B}\left(\hbar\kappa\lambda^{2}\right).\label{UR2}
\end{equation}
The above result should prove useful in studies devoted to quantum aspects of EPR-based ghost imaging
\cite{EPRghost} and security protocols for compressive quantum imaging \cite{compresive2, compresive}. A "periodic" variant of Eq. (\ref{UR2})  (when $M(x)$ is a periodic function) is already applied in experimental entanglement detection with periodic amplitude masks \cite{OurMasks}.

The derivation of (\ref{UR2}) is very simple. Due
to Parseval's theorem  from Fourier analysis, the left hand
side translates directly to the $\lambda$-domain. To obtain the
right hand side we apply the UR for characteristic
functions from Theorem 1, with $\lambda_{x}=\lambda$ and $\lambda_{p}=\kappa\lambda$.  It is worth mentioning that Eq. \ref{UR2} remains valid for any complex--valued function $M(x)$.

\paragraph{ChUR and various theories in physics---}Besides its fundamental interest, the ChUR derived in this paper is related to several issues across various fields of physics. 
First of all, our approach remains valid with $\Phi(\lambda_{x})$ and $\tilde\Phi(\lambda_{p})$ substituted by $\left\langle \Psi\right|U\left|\Psi\right\rangle$ 
  and $\left\langle \Psi\right|W\left|\Psi\right\rangle$, whenever the unitary matrices $U$ and $W$ satisfy the Weyl-type commutation relations $UW\!=\!e^{i \phi} WU$. A prominent example provided by Schwinger \cite{Schwinger} and given by $\phi=2\pi/d$,  where $d$ is the dimension of the Hilbert space, is a sort of prerequisite for the fruitful theory of Mutually Unbiased Bases.

Moreover, since Theorem 1 involves operators of the form $e^{i\lambda\hat{O}}$ it becomes valuable when the operator $\hat{O}$ does not exist itself (consequences of so called
Stone\textendash von Neumann theorem). A particularly interesting example of the number-phase uncertainty \cite{CN, Luis2, Shepard} (phase operators are not well defined) has just been described \cite{Luis} along the lines of Theorem 1. 
Here we would like to briefly touch upon the broad theory of Loop Quantum Gravity \footnote{We shall restrict ourselves to mention only few aspects of the whole theory, those which are interesting from the uncertainty relations point of view. A reader interested in Loop Quantum Gravity is encouraged to see \cite{RovelliBook}.}, in which the so called Ashtekar connection \cite{AshtekarOld} $A_a^i(x)$ \footnote{The lower index "$a$" refers to spatial coordinates, while the upper (so called internal) index "$i$" is related to $SU(2)$ gauge group.} plays the role of the canonical "position" variable in a field-theoretical sense \footnote{The canonical momentum $E^a_i(x)$ is the so called Ashtekar electric field.}. While moving to a quantum description, the problem appears as there is no local operator $\hat A_a^i(x)$, and one needs to resort to unitary holonomies. The standard approach to quantum uncertainty relations cannot thus be directly applied, however one can involve the Weyl algebra \cite{WeylLQG} and  use Theorem~1.

To explain better the idea behind the above prescription, we would like to discuss a very simple case from the field of Loop Quantum Cosmology.  To this end we start with the well known FLRW metric
\begin{equation}
ds^{2}=-c^{2}dt^{2}+a^{2}(t)d\Sigma^{2},
\end{equation}
where $a(t)$ denotes the dimensionless scale factor and $\Sigma$ refers to the 3-dimensional space. We further recall two time-dependent variables \cite{Ashtekar2}: $b=\dot{a}/a$ and $V=V_0 a^{3}$, denoting the Hubble parameter and the physical volume of the expanding Universe respectively ($V_0$ is the coordinate volume). These variables satisfy the following Poisson bracket relation \cite{Ashtekar2}
\begin{equation}
\left\{ b,V\right\} =\pm4\pi G/c^{2}\equiv Q,\label{poisson}
\end{equation}
where the $\pm$ sign depends on the orientation and is irrelevant in our considerations. If the operator $\hat b$ existed, then (\ref{poisson}) would lead us  to the UR: $\sigma_b\sigma_V\geq 4\pi\hbar G/c^2$ as stated \footnote{Note that Eq. 11.21 of \cite{RovelliBook} uses a bit different variables.} in Eq. 11.21 of \cite{RovelliBook}. Since the Ashtekar connection of this well-studied model is given by $A_a^i=\beta \dot{a}\delta_a^i$ \footnote{$\beta$ denotes the Barbero-Immirzi parameter, whose exact value has no relevance for our example as long as $\beta\neq0$.}, it becomes obvious that $b$ cannot be promoted to a quantum mechanical operator. As this limitation is not shared by the holonomy $U_b(\lambda_b)=e^{i \lambda_b b}$, one can use $U_b$ together with  $e^{i \lambda_V \hat V}$ and apply Theorem 1. In particular, if we set $\lambda_V=\pi/(\hbar Q\lambda_b)$, so that the bound $\mathcal{B}$ is equal to $1$, and use (\ref{lowerSD}) to extract the variance $\sigma_V^2$, the ChUR provides the uncertainty relation of the form:  
\begin{equation}
\frac{4\hbar G}{c^{2}}\lambda_{b}\left|\left\langle U_b(\lambda_{b})\right\rangle \right|\leq\sigma_{V}.
\end{equation}
The above UR is a formally right way of bounding the fluctuations of the volume of the Universe in terms of the volume shift operator $U_b$, relevant for understanding of the big-bang singularity \cite{Bojowald}. Note also  that this example actually represents Quantum Mechanics subject to the Bohr compactification. In other words, Theorem 1 is the only path towards URs in theories (such as Loop Quantum Cosmology \cite{Fewster}) with the Bohr compactification involved.

Looking into future, a further development of the discrete counterpart of the presented theory might bring useful results, for instance, in compressed sensing, as the Dirac comb state has no counterpart in various discrete systems.  Generalizations of URs for the electromagnetic field (as discussed in \cite{Ashtekar1}) might bring a better physical insight into the role played by Gauss linking numbers, or contribute to a better understanding of quantum effects for the gravitational field in a hot universe \cite{IBBPRD}. We also believe that our approach will be influential to the theory of
quantum optical characteristic functions. Questions about non-classicality of light are being asked and studied  \cite{Agudelo1} in terms of the characteristic $P$-function reconstructed from the data accessible in experiments \cite{Agudelo2}, with a relevant filtering procedure based on auto-correlations \cite{Agudelo3}.

\acknowledgments 
We would like to thank Alfredo Luis for fruitful discussions and correspondence. \L R acknowledges financial support by the grant number 2014/13/D/ST2/01886
of the National Science Center, Poland. Research in Freiburg is supported
by the Excellence Initiative of the German Federal and State Governments
(Grant ZUK 43), the Research Innovation Fund of the University of
Freiburg, the ARO under contracts W911NF-14-1-0098 and W911NF-14-1-0133
(Quantum Characterization, Verification, and Validation), and the
DFG (GR 4334/1-1). DST and SPW acknowledge financial support from the Brazilian agencies CNPq, CAPES, FAPERJ and the Instituto Nacional de Ci\^encia e Tecnologia - Informa\c{c}\~ao Qu\^antica.


\end{document}